# Landauer resistivity dipole at one dimensional defect revealed via near-field photocurrent nanoscopy

Francesca Falorsi[1], Marco Dembecki[2], Christian Eckel[1], Monica Kolek Martinez de Azagra[1], Kenji Watanabe[3], Takashi Taniguchi[4], Martin Statz[1], R. Thomas Weitz[1,5,*]

[1] 1st Institute of Physics, Faculty of Physics, Georg-August-University Göttingen, Göttingen
[2] Physics of Nanosystems, Faculty of Physics, LMU Munich, Germany; current address: WSI, TUM, Garching, Germany
[3] Research Center for Electronic and Optical Materials, National Institute for Materials Science, 1-1 Namiki, Tsukuba 305-0044, Japan
[4] Research Center for Materials Nanoarchitectonics, National Institute for Materials Science, 1-1 Namiki, Tsukuba 305-0044, Japan
[5] International Center for Advanced Study of Energy Conversion, Göttingen ICASEC
* corresponding author: thomas.weitz@uni-goettingen.de

The fundamental question how to describe Ohmic resistance at the nanoscale has been answered by Landauer in his seminal picture of the so-called Landauer resistivity dipole. This picture has been theoretically well understood, however experimentally there are only few studies due to the need for a non-invasive local probe. Here we use the nanometer lateral resolution of near-field photocurrent imaging to thoroughly characterize a buried monolayer – bilayer graphene interface as an ideal one dimensional defect for the Landauer resistivity dipole. Via systematic tuning of the overall charge carrier density and the current flow we are able to detect the formation of Landauer resistivity dipoles due to charge carrier accumulation around the one dimensional defects. We found that, for Fermi energy values near the charge neutrality point (i.e. at low hole or electron doping), the photocurrent exhibits the same polarity as the applied source-drain voltage, which is consistent with changes in carrier concentration induced by the Landauer resistivity dipoles. This signature is no longer evident at higher charge carrier density in agreement with the performed numerical calculations. Photocurrent nanoscopy can thus serve as non-invasive technique to study local dissipation at hidden interfaces.

As the scaling of electrical integrated circuits down to a few nanometer sizes continues, the associated increase in current densities and the concomitantly growing impact of detrimental effects such as electromigration and excess heat generation become increasingly problematic [1]. Consequently, a better understanding of the precise mechanisms that govern electronic flow and resistivity on a nanoscopic level is a topic that continues to attract experimental and theoretical research [2–5]. One fundamental nanoscopic mechanism of resistance formation at the nanoscale, proposed by Landauer, involves the creation of residual resistivity dipoles, often referred to as Landauer resistivity dipoles (LRD)[6–9]. According to the LRD theory, a localized dipole forms around a defect during current flow, serving as a nanoscopic cause of resistance. Consequently, the voltage drop throughout the material is not homogeneous but rather highly concentrated around these defects.

To add further understanding to this picture, it would be helpful to be able to directly observe this dipole. The spatial long-range dipolar electric field generated by the LRD is however challenging to access experimentally in three dimensions due to its inherently low dipole strength [10]. It was recognized early on [9] that in two dimensions, the spatial extent of the LRD should be larger than in 3D, due to decreased screening. Korenblum and Rashba later specified that local inhomogeneities of the conductivity even can have a global impact on the current distribution [10]. In general, the local potential build up by the LRD in 2D is given by $V(r) \sim p \cdot cos(\theta)/r$, where $p$ is the dipole moment, $\theta$ the angle between $r$ (the distance from the scatterer) and the current[9]. In the diffusive limit, the total electric field caused by the dipoles surrounding the scatterer, $E^{dip}=n_s \cdot p$, is equal to the Ohmic electric field in the sample. Consequently, the LRD dipole strength is given by $p \sim j/(\sigma n_s)$, with $\sigma$ the conductivity of the sample, $j$ the incident current density and $n_s$ the density of scatterers [7]. The potential build-up due to the LRD increases with current density or decreasing conductivity of the sample. Especially interesting are thus experimental 2D systems, such as graphene and other van der Waals materials [11–14], in which both $\sigma$ and j can be tuned in-situ, making them an ideal system to study the LRD. Graphene, in particular, one of the most extensively researched 2D materials [15–17], serves as an excellent platform to investigate the effects of scatterer on a local scale.

Indeed, the localized voltage drop across localized defects in graphene and ML/BL graphene interface junctions has been experimentally measured on graphene sheets grown on silicon carbide (SiC) using scanning tunneling potentiometry (STP) at low temperatures (ranging from 6K to 80K) [2,18–21] and conductive AFM [22]. In all of these studies the Fermi energy was maintained constant. Ji et al. [21] and Willke et al. [20] showed the current dependence of the localized voltage drop, which has a sign and magnitude related to the current flow. STP has also been employed to measure the localized voltage drop induced by the current flow in p-n junctions of graphene on h-BN [4] and on localized charged defects in graphene [23]. In this last work, the expected dependence of the voltage on the current density was also studied. While fundamental studies show part of the expected response of the LRD at cryogenic temperatures, signatures of LRD at technologically relevant conditions (room temperature, buried interfaces) or as function of current under control of the Fermi energy (and with it the conductivity) have to the best of our knowledge not been performed.

Here we discuss systematic investigations of a graphene monolayer to Bernal-stacked bilayer interface (ML/BL interface) as function of electrostatically tuned charge carrier density (which controls $\sigma$) and current density *j* with the goal to locally identify the LRD. The wavefunction mismatch at this interface reduces electron transmission, allowing us to consider it as a reflective wall [20]. To study the local buildup of a density difference at the ML/BL interface as function of $\sigma$ and *j*, we use photocurrent imaging, which has been successfully used to study lateral charge accumulation in van der Waals materials such as graphene [24]. Governed by the photo-thermoelectric effect in graphene, this method is particularly sensitive to the local Seebeck coefficient and thus chemical potential [25–27]. To allow for a nanoscale local resolution, we use near-field optical microscopy (SNOM) [28–30] which enables high-resolution mapping of photocurrent at a scale equivalent to the tip radius (~20 nm) [31,32]. Furthermore, it enables us to acquire photocurrent images of h-BN protected (i.e. buried) ML/BL interfaces. High-resolution SNOM photocurrent nanoscopy in graphene systems has been used for the detection of propagating collective carrier modes[33,34], domain walls [35–37], and electronic transport phases in twisted graphene samples at low temperatures [38,39].

We exploit the sensitivity of the photocurrent measured with SNOM to the local chemical potential gradient and use it to study the formation LRD at ML/BL graphene interfaces during current flow at ambient conditions. We complement our experimental results with numerical calculations, which confirm that the formation of Landauer resistivity dipoles at the interface leads to a change in polarity of the photocurrent response upon changing the applied bias when the device is near its charge neutrality point. This signature becomes less evident at higher doping concentrations, where the device response is dominated by carrier type, with minimal influence from the LRDs. With our measurements we are thus firstly able to consistently map the density and current dependence of the LRD in real-space.

We first explain the general mechanism by which a photocurrent ($I_{PC}$) in graphene can emerge, before introducing our measurements. When electrons in graphene are excited by incident radiation, the absorbed photon energy is rapidly transformed into the formation of a hot Fermi distribution by electron-electron scattering. Due to the low electron-phonon coupling, electron cooling is inefficient, allowing hot electrons to remain decoupled from the lattice for relatively long times (from ps up to ms)[26,27,40,41]. The hot electrons can then produce a net photocurrent at local regions of different doping or potential due to their different thermopower (i.e. Seebeck coefficient). In our specific case of the ML/BL interface, when we consider x the direction perpendicular to the interface, $I_{PC}$ can be expressed as [32]:

$$I_{PC} = -\frac{1}{L}\int_0^L \sigma(x)S(x)\frac{\partial T}{\partial x}dx \tag{1}$$

Where $\sigma(x)$ is the spatially dependent conductivity, $T(x)$ is the elevated electron temperature profile induced by the SNOM tip and $S(x)$ is the spatially dependent Seebeck coefficient and L is the distance between the source and drain contacts. A spatially inhomogeneous Seebeck leads to a non-negligible photocurrent. In our case, if we consider the only source for a spatially varying Seebeck coefficient the ML/BL interface, the Seebeck coefficient can be modelled as a step function, and the photocurrent signal depends on the Seebeck coefficient difference across the interface $\Delta S = S_{BL} - S_{ML}$ [42].

In the semiclassical Boltzmann formalism under the relaxation time approximation the Seebeck coefficient can be written as [43]:

$$S = -\frac{1}{eT}\frac{\int (\varepsilon-\mu)\frac{\partial f}{\partial \varepsilon}\sigma(\varepsilon)}{\int \frac{\partial f}{\partial \varepsilon}\sigma(\varepsilon)} \quad \text{with} \quad \sigma(\varepsilon) = e^2 v(\varepsilon)^2 DoS(\varepsilon)\frac{\tau(\varepsilon)}{2} \tag{2}$$

Where e is the electron charge, $T$ the temperature, $f$ is the Fermi distribution, $\sigma(\varepsilon)$ is the energy dependent conductivity and $\mu$ is the chemical potential. The electrical conductivity $\sigma$ depends on the electron velocity $v$, the density of states $DoS$ and the scattering time $\tau$ [43,44]. Thus, the Seebeck coefficient serves as a measure of the local chemical potential and, consequently, the local carrier density, as these two quantities are directly related [16,45]. This implies that the scanning photocurrent measurements should act as sensitive probe for the local occurrence of the LRD.

To perform local mapping of LRD, photocurrent measurements are performed using an SNOM with an illumination wavelength of λ=10.551 µm. Integrated electrical measurement units allow parallel to the optical also photocurrent measurements and in-situ control of the source-drain ($V_{SD}$) and back-gate ($V_{GS}$) voltages applied to the system (Figure 1a). The locally detected photocurrent signal is demodulated with the first or second harmonic of the tip oscillation frequency Ω, allowing the suppression of the far-field background signal.

To analyze the carrier accumulation of impinging electrons at the ML/BL interface, the two main contacts for the source and drain are placed parallel to the interface. With this geometry, it is possible to probe the Seebeck coefficient gradient perpendicular to the current flow induced by the LRD [37,46]. In both samples analyzed for this study, the graphene interface is encapsulated between two hexagonal boron nitride (hBN) flakes, the upper one being thinner than 7 nm, to retrieve the photocurrent signal originating from the underlying graphene [31]. Graphite gates are used to improve signal quality and reduce energetic disorder in the active layers. A schematic of the investigated device structure is shown in Figure 1b, while AFM topography scans of the two samples considered for this study are shown in Figure 1c and S2. Since the samples show similar results, the discussion below focuses on the sample shown in Figure 1c, while the complementary results of the second sample are shown in the Supplementary Information.

Figures 1d and 1e show a spatial scan of the area highlighted by the dotted red rectangle in Figure 1c performed with the SNOM, Figure 1d shows the 3$^{rd}$ harmonic optical amplitude image, while Figure 1e shows the 2$^{nd}$ harmonic PC signal that is recorded simultaneously. The schematic next to Figure 1e highlights that in the acquisition of this photocurrent signal, none of the other parameters that could be varied during the measurement, such as the gate-source or source-drain voltage, are adjusted; instead, the contacts are left floating. This scheme is maintained throughout the manuscript, with the main varying parameter of each respective photocurrent scan highlighted in red.

The photocurrent scan in Figure 1e shows, that in the case of vanishing $V_{DS}$ the main signal is generated by the spatial inhomogeneities of the samples, such as cracks or bubbles that are created during fabrication. In these regions, the carrier density and strain of the sample varies, resulting in a spatial texture of the Seebeck coefficient across the sample[47,48]. In the rest of the work we will focus on the photocurrent traces recorded perpendicular to the ML/BL interface indicated by the faint gray line in the Figure 1d. We use the interface as a well-localized one-dimensional (1D) defect for the local investigation of the LRD and to this end systematically vary $V_{SD}$ and $V_{GS}$. To establish the photocurrent measurements, we map (Figure 2a) the photocurrent as function of charge carrier density which is electrostatically controlled by the graphite gate ($V_{GS}$ voltage). An additional small constant $V_{SD}$ (1 mV) is applied to monitor the system resistance, which is shown for each point of the scan in the curve to the left of the scan. The dotted line indicates the position of the ML/BL interface. This position is obtained by aligning the photocurrent map with the 3$^{rd}$ harmonic optical amplitude images, where the ML/BL interface is discernable by a difference in optical contrast, as shown for example in Supplementary Figures S5 and S6. This alignment method is used throughout the manuscript.

As the gate voltage is tuned from negative to positive values, the induced carrier density transitions from holes to electrons, as shown in the schematic representation in Figure 2a. In this density region, the photocurrent near the interface exhibits a double sign switch, as highlighted by the line cut in Figure 2b, which is taken directly at the ML/BL interface. This behavior is related to the thermoelectric origin of the photocurrent, which depends on the difference in Seebeck coefficients between the monolayer and the bilayer graphene $\Delta S$ [42]. To understand the signal in more detail we have numerically calculated the Seebeck coefficient of the ML, the BL and the respective $\Delta S$ in Figure 2c. The Seebeck coefficient is simulated based on formula 2, without considering the density dependence of $\tau$ [43,49,50], more details are provided in Supplementary Section S1. The simulated $\Delta S$ (Figure 2c) shows, as expected, the same density dependence as the photocurrent, with a double sign switch. The reason is, that the difference in band structure between monolayer and bilayer graphene leads to a smoother transition of the Seebeck coefficient of the bilayer from electron- to hole-dominated transport (or from negative to positive Seebeck coefficient) compared to the monolayer [51,52], leading to the double sign switch observed in the photocurrent response. The consistency of electrical measurements and simulations confirms the good understanding of the system.

The photocurrent signal has also a characteristic extend in the direction perpendicular to the interface (which is not included in the above calculations). In Figure 2a one can see, that after a maximum value around the CNP, the photocurrent in the ML decreases, and finally disappears for $V_{GS}$> 0.9 V on the electron and <-0.5V on the hole side. In contrast, the photocurrent in the bilayer remains significant in the entire gate voltage window. The different distance dependence of $I_{PC}$ in the ML vs. the BL is related to the fact that the $I_{PC}$ signal does not only depend on ΔS, but is convoluted with the spatial extend of the temperature profile induced by the local SNOM illumination (see equation (1)). This profile is determined by the "cooling length" [27,53]. The cooling length of the hot electrons is defined as $L_H = \sqrt{\frac{\kappa}{g}}$, where κ is the thermal conductivity and $g$ is a constant that accounts for the dispersion of the heat into the substrate [32]. According to the Wiedemann-Franz law, κ is proportional to the conductivity of the sample [54]. Indeed, a longer cooling length can be assumed for the BL since it has a lower resistance than the ML (as shown in the measurements in S4).

After having characterized the ML/BL in detail, we turn to the investigation of the LRD. The LRD potential V(r)~j/(σ·r) at the ML/BL interface translates into a local carrier buildup and consequently to a ΔS, which we can locally measure as $I_{PC}$. To investigate the local carrier accumulation at the interface due to the formation of LRD, the photocurrent response of the ML/BL interface is studied with respect to the applied bias. To this end a line scan along the same position as in Fig. 2a is performed but with varying $I_D$ at a fixed $V_{GS}$. Figure 3a shows these scans for three constant $V_{GS}$, in the hole doped regime ($V_{GS}$ =-100 mV), at the local CNP ($V_{GS}$ = 0 V) and in the electron doped regime ($V_{GS}$ = 150 mV). We examine specifically the low-bias regime ($|V_{SD}|$ < 30 mV), where $I_{PC}$ is generated by the PTE and per-se is not expected to be dependent on the applied $V_{SD}$ or $I_D$. This is indeed also seen for scans taken in the electron and hole doped regime at $V_{GS}$=-100 mV and $V_{GS}$ = 150 mV (Figure 3a). We note in passing that we can consequently also rule out that $I_{PC}$ is due to the bolometric effect where $I_{PC}$ should indeed depend on magnitude and sign of applied bias [55–57] (see also Figure S6 for more details).

In contrast to these scans at high carrier density, a strong dependence of $I_{PC}$ on the $I_D$ flowing through the ML/BL interface is visible for the scan taken around the CNP (Figure 3a). We attribute this to the local formation of a LRD – due to the applied $I_D$ – which leads to a density difference $\Delta n_0$ across the interface and with it a $\Delta S$ which we detect as localized $I_{PC}$. To corroborate that a charge carrier density difference $\Delta n_0$ at the ML/BL interface leads to a $\Delta S$ and $I_{PC}$ as we measure it, we have performed calculations (Figure 3b) of $\Delta S$ at the ML/BL junction as function of $\Delta n_0$ (which is caused by LRD V(r) ~$I_D$) and n (which is controlled by $V_{BG}$). In our calculation we add for each datapoint $\Delta n_0/2$ carriers to the ML and -$\Delta n_0/2$ to the BL. Most interesting is the region within the dashed box (Figure 3b) near the charge neutrality point where a sign change of $I_D$ also leads to a sign change of $\Delta S$, which similarly should be present in $I_{PC}$. This result is expected, because near the CNP, a small $\Delta n_0$ (induced by $I_D$ and the LRD) is sufficient to induce electron transport on one side of the interface and hole transport on the other, leading to large differences in the Seebeck coefficients of the two regions (in this region of small n, $d\Delta S/dn$ is also very large, see Figure 2c). At higher n, the overall impact of the LRD induced $\Delta n_0$ on $\Delta S$ is comparatively small.

The measured $I_D$ dependence of $I_{PC}$ is indeed consistent with our calculations under the hypothesis of localized LRD formation at the ML/BL interface: a reversal in the applied $I_D$ sign results in a change in the $I_{PC}$ polarity at the CNP, while $I_{PC}$ is independent of $I_D$ at higher charge carrier densities.

We have also analyzed the detailed $I_{PC}$ (vs. $I_D$) dependence in a small $V_{GS}$ window around the CNP in Figure 4. There, one can discern that $I_{PC}$ is characterized by two main features: it is symmetric around the interface, with an intensity decreasing with the increasing distance from the interface and shows a symmetric increase around the interface as function of $I_D$. There is also a systematic $V_{GS}$ dependence of $I_{PC}$. $I_{PC}$ increases as function of doping (the CNP in this measurement is around $V_{GS}$=-10 mV). To understand this feature in more detail, vertical line cuts taken in the monolayer at the positions indicated by the three arrows at the top of Figure 4a are shown in Figure 4b. These cuts compare the $I_{SD}$ dependence of the $I_{PC}$ at a fixed x-position in the ML right at the ML/BL interface at different n, and highlight that the $I_D$ for which the recorded photocurrent switches sign (i.e. the photocurrent values cross the $I_{PC}$=0 line) depends on $V_{GS}$. To analyze this behavior, a simulation of spatial dependence of the photovoltage generated by a SNOM tip with respect to different $\Delta n_0$ values was performed (Figure 4c, for details see SI). The simulated spatial scans are calculated for two different charge carrier densities of the system, n=0 cm$^{-2}$ and n=0.4·10$^{-12}$ cm$^{-2}$: The simulated scans in Figure 4c are thus comparable to the ones shown in Figure 4a. Both for the simulations and measurements, the value for which the photocurrent switches sign shifts with the value of the total charge carrier density. While at n=0 cm$^{-2}$, the sign switch occurs precisely at $\Delta n_0$ =0 cm$^{-2}$, at n= 0.4·10$^{12}$ cm$^{-2}$, a strong negative shift of $\Delta n_0$ =-0.8·10$^{12}$ cm$^{-2}$ is necessary to observe the sign switch of the photocurrent. In the experiment, the $\Delta n_0$ accumulation at the BL/ML interface is a consequence of the LRD. Finally, the increase in $I_{PC}$ with increasing $\Delta n_0$ (or $I_D$) is also consistent with the LRD picture where V(r)~I.

In conclusion, the nanoscopic photocurrent detection based on a SNOM at ambient conditions allows the observation of Landauer resistivity dipoles in the electronic current flow and directly visualize the charge carrier accumulation at a mono-bilayer graphene interface, which due to the local wave-function mismatch acts like a localized 1D defect. The possibility to analyze local, buried defects with a non-invasive method could be of use for the analysis of integrated circuits where local resistances could be detected prior to device failure.

# Experimental methods

**Device Fabrication:**

The stacks were fabricated using the dry transfer method described in reference [58]. The graphene and hBN flakes were obtained through mechanical exfoliation, respectively with scotch tape (*Scotch, Magic tape*) and Blue Nitto tape (*Nitto inc., SWT20+*) onto a Silicon/Silicon dioxide (300 nm) substrate. The graphene flakes were exfoliated either from natural graphite crystals (*NGS trading and consulting*) or a block of highly ordered pyrolytic graphite (*Momentive Performance Materials Inc.*). The hBN flakes were grown by K. Watanabe and T. Taniguchi. The flakes were first initially selected with an optical microscope in bright-field mode (*ZEISS,* model Axio Scope.A1).

The graphene layer thickness is confirmed via Raman spectroscopy. Two different setups were used: a commercially available LabRam HR Evolution (*Horiba*) and a Raman assembly composed of a microscope 100x objective (MPlanFL N 100*, Olympus*) and a Spectrometer iHR550 (*Horiba Scientific*) with a 1800 lines per mm grating. Both are coupled to a 532 nm wavelength laser (torus 532*, Laser Quantum*).

The topography of the flakes of the stack is checked with atomic force microscopy (AFM) before stamping. Three different setups were used: an Asylum Jupiter AFM by *Oxford Instruments*, and Dimension 3100 and Dimension Icon (*Bruker)*, all operated with Tap300Al-G (*NanonAndMore*) tips in tapping mode.

An electron beam lithography microscope (*Raith*) operated with an accelerating voltage of 10 kV was used to pattern the electrical contacts. For small contacts a dose of 110 $\mu$C cm$^{-2}$ with a 7.5 $\mu$m aperture was utilized, while for bigger contacts and pads a dose of 170 $\mu$C cm$^{-2}$ with a 60 $\mu$m aperture was used. The layer of resist for the e-beam procedure was obtained following the procedure described in reference [59]. Finally, an 2 nm adhesion layer chromium (with a rate of around 0.43 Å/s) and 60 nm gold contacts (with a rate of around 0.9 Å/s) are evaporated via either thermal evaporation (evaporation chamber from *BesTec*) at pressures of around 10$^{-6}$ mbar or electron-beam physical vapor deposition chamber (electron-beam PVD) at pressures of less than 5 x10$^{-7}$ mbar. For one of the two samples studied 1D etch contacts were created by etching through the top hBN flake. For this purpose an inductive coupled plasma-reactive ion etching (ICP-RIE) device (*Oxford Instruments*, model ICP-RIE Plasmalab System 100 ) was used. To etch mainly hBN, a mixture of SF6 and Ar was utilized (with a flow rate of 10/5 sccm, an ICP power of 70 W and an RF power of 50 W for an etch rate of 9 nm min$^{-1}$), while for graphene O2 was utilized (with a flow rate of 10 sccm, an ICP power of 40 W, an RF power of 150 W for an etch rate of 7 nm min$^{-1}$).

The samples are then glued on a home build chip carrier with a silver conductive paste. The chip carrier is formed by an insulating plastic and contains 18 gold conductive pins, in order to allow electrical measurements at the SNOM setup. The gold pads are then bonded with a wedge bonder (MEI 1204W, *Marpet Enterprises or* K&S 4500 Series, *Kulicke & Soffa Ltd*) to the gold pins of the chip carrier.

**Photocurrent measurements:**

The near field scattering microscope and PC images are taken with a commercial s-SNOM (from *Neaspec Company*) coupled to a tunable $CO_2$ laser (from Access laser, model L4G) with wavelengths of 9.2–10.78 µm. The SNOM is based on an atomic force microscopy (AFM) operated in tapping mode with a tapping amplitude of Δ$z$=90 nm. All the images were taken with two different tips: Arrow-NCPt and Arrow-NCx76-50 (*Nanoworld*), both characterized by a tapping frequency Ω of ~270 kHz. The power of the laser during the measurement was set to around 10 mW.

In order to perform electrical measurements, the s-SNOM is connected to two source measurement units (Keithley 2450, *Tektronix*) and to a current amplifier (DHPCA-100, *FEMTO*) which converts the signal into a voltage and amplifies the AC part of the signal. The output of the amplifier is then transmitted to the DAQ card of the SNOM and the signal is demodulated with the Ω frequency of the tip with an internal lock-in amplifier. For each pixel of a scan, together with the optical signal, two additional values will be recorded, an amplitude $A_{PC}$ and a phase $P_{PC}$ in rad, for all the different harmonics of the tip oscillation frequency Ω. These values are respectively connected to the strength and the direction of the local photocurrent in the system. The analysis of the PC derived from the higher harmonics allows for the isolation of the PC signal developed from near-field interaction between the tip and the sample from the background signal, dependent on unfocused light, ensuring the nm resolution of the obtained signal. The second harmonic of the tip oscillation frequency is primarily analyzed for the sample shown in the main manuscript, while first-order harmonic is used for the sample presented in the SI, with the general conclusions being the same. In the second sample the higher harmonic could not be used as its signal was too small.

**Author contributions**

F.F., M.D. and R.T.W. conceived the research. F.F and M.D. fabricated the devices and performed the photocurrent measurements with the help of C.E. and M.K.M.d.A. F.F. & M.D. performed the thermoelectric transport simulations, supervised by M.S.. K.W. and T.T. grew the h-BN crystals. All authors discussed and interpreted the data. R.T.W. supervised the experiments and the analysis. The manuscript was prepared by F.F. and R.T.W. with input from all authors.

**Acknowledgements**

We acknowledge discussions with Leonid Levitov and Martin Wenderoth. Fritz Keilmann is acknowledged for discussions and experimental help in the initial phases of the project. K.W. and T.T. acknowledge support from the JSPS KAKENHI (Grant Numbers 21H05233 and 23H02052), the CREST (JPMJCR24A5), JST and World Premier International Research Center Initiative (WPI), MEXT, Japan. R.T.W. acknowledges partial funding from the Center for Nanoscience (CeNS) and the Nanosystems Initiative Munich (NIM). This project has been funded by the Deutsche Forschungsgemeinschaft within the Priority Program SPP 2244 "2DMP". We acknowledge funding from the SFB 1073, project B10 and A05.

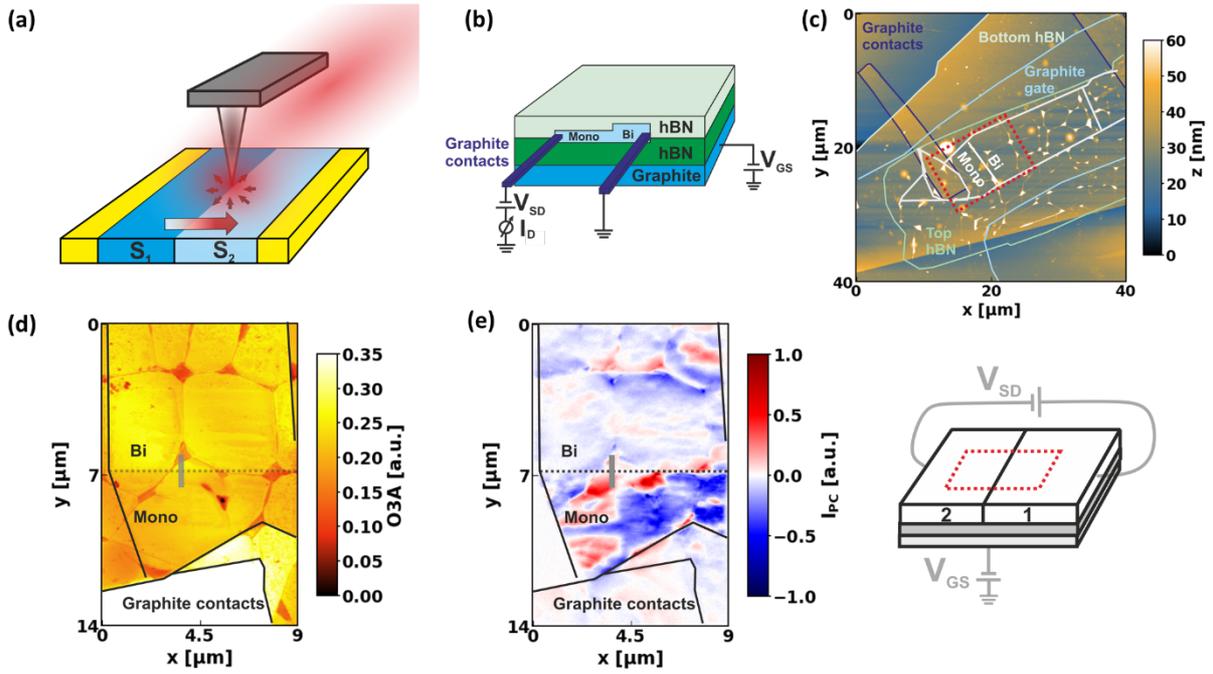

*Figure 1: (a)* Schematic representation of the setup used to analyze the spatial dependence of the photocurrent. *(b)* Schematic representation of the sample geometry. *(c)* AFM image of the sample studied with the different layers outlined. *(d)* 3$^{rd}$ harmonic optical amplitude of area highlighted by the red rectangle in *c*. *(e)* 2$^{nd}$ harmonic photocurrent image, recorded simultaneously with the image in *d*. The schematic on the right highlights the fact that during the acquisition of the photocurrent map neither $V_{GS}$ nor $V_{SD}$ are varied.

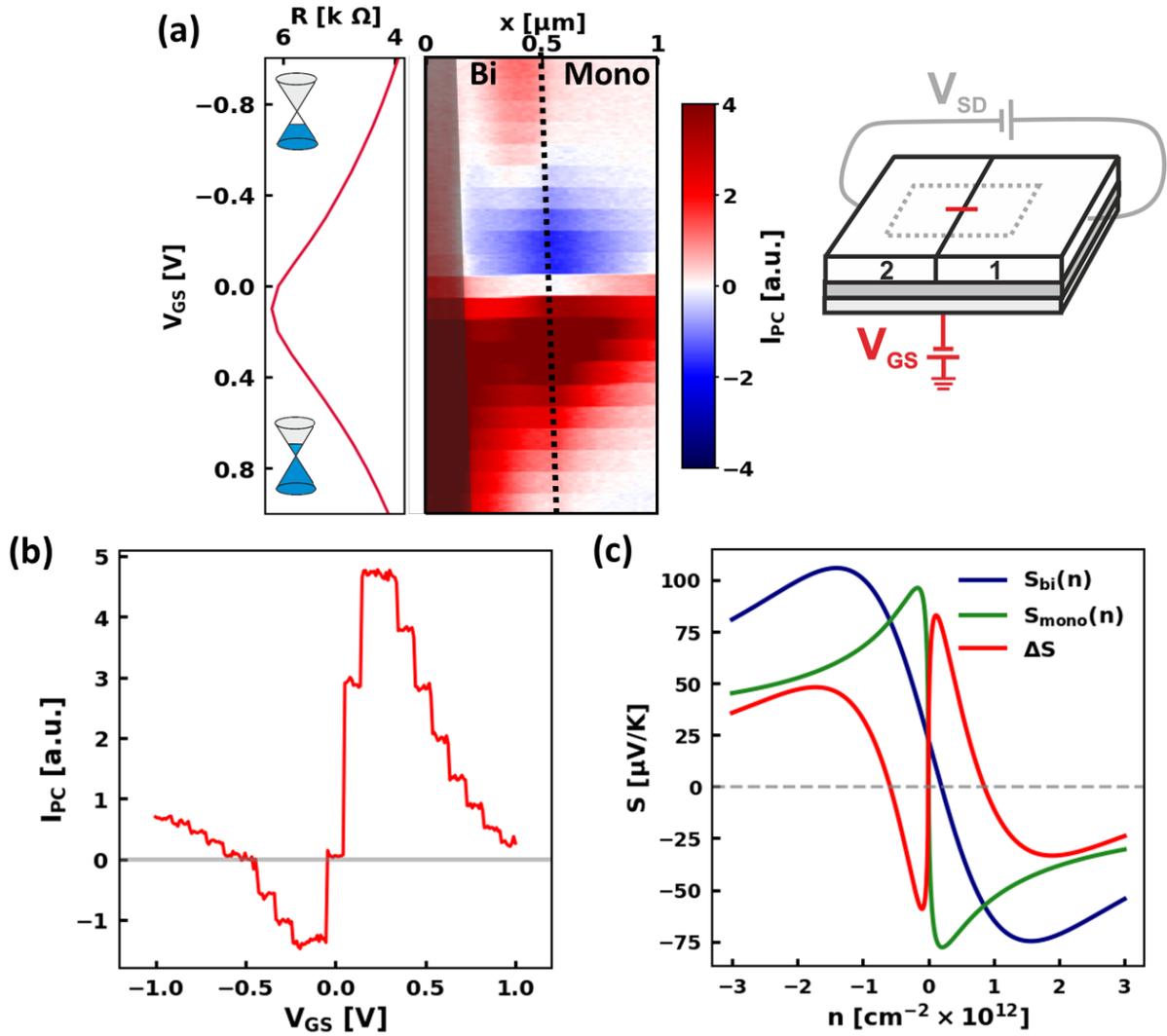

***Figure 2***: *(a)* 2$^{nd}$ *harmonic* $I_{PC}$ *measured along a line across the ML/BL interface as a function of* $V_{GS}$. *The* $V_{GS}$ *varies from -1 V to 1 V in steps of 0.1 V, with 10 line traces recorded for each step. The interface is marked by the black dotted line while under the gray shaded area on the left of the map a topographical defect is present. A* $V_{SD}$ = 1 *mV is applied to simultaneously monitor the resistance (graph to the left of the photocurrent scan). From the top to the bottom of this map the transport changes from a hole to an electron dominated transport, as schematically sketched with the Fermi level of the Dirac cones in the resistance plot. The schematic on the right highlights the fact that during the acquisition of this scan* $V_{GS}$ *is varied while* $V_{SD}$ *is maintained constant.* **(b)** *Vertical line cut of the photocurrent signal taken at ML/BL interface.* **(c)** *Numerically calculated Seebeck coefficient of the ML, in green, and the BL in blue with respect to the charge carrier density (at 300K). The difference between the two Seebeck coefficients is shown in red.*

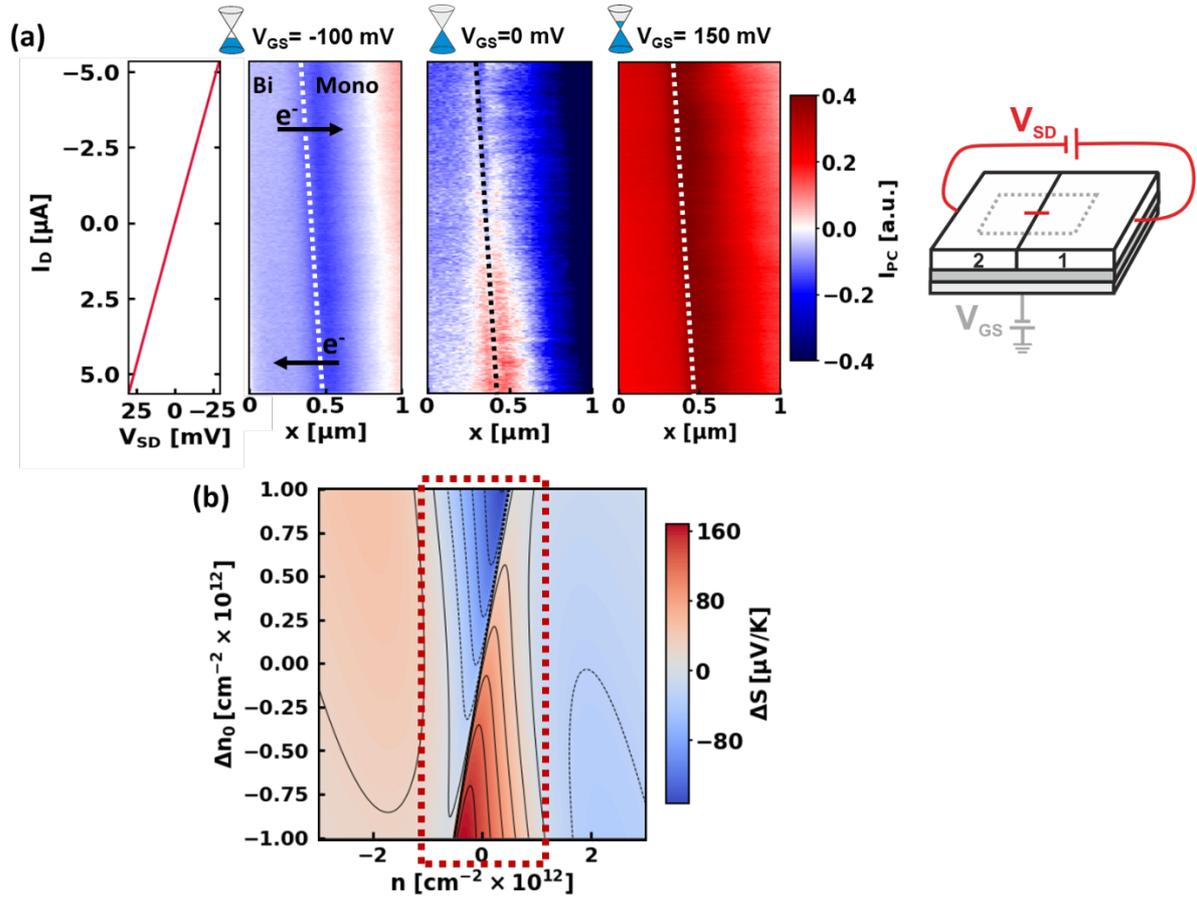

*Figure 3 (a)* 2nd harmonic photocurrent maps measured along a line across the ML/BL interface (highlighted in Figure 1c), as a function of $I_D$. The data are acquired with $V_{SD}$ ranging from -30 mV to 30 mV in steps of 2 mV, with 10 lines measured for each step. The three maps are measured for three different $V_{GS}$ values, as indicated by the title on top of each map, $V_{GS}$=-100 mV, $V_{GS}$=0 mV and $V_{GS}$=150 mV. The graph to the left of the maps shows the $I_{SD}$ with respect to the $V_{SD}$ applied, recorded during the photocurrent map taken at $V_{GS}$= -100 mV; y- axis and color bar are shared. The interface is indicated with the dotted lines. The arrows indicate the direction of electron flows respectively at the top and bottom halves of the maps, corresponding to the sign of the $V_{SD}$. The schematic on the left highlights that the $V_{SD}$ is varied in the acquisition of these scans, while the $V_{GS}$ is maintained constant. *(b)* Simulated **d**ifference of the Seebeck coefficient value $\Delta S$ between the monolayer and the bilayer graphene with respect to the total carrier density n, controlled by $V_{GS}$ and an additional parameter $\Delta n_0$, independent of n and mimicking the role of the deviation of the carrier density n induced by the LRD around the interface which is a consequence of the applied $I_D$.

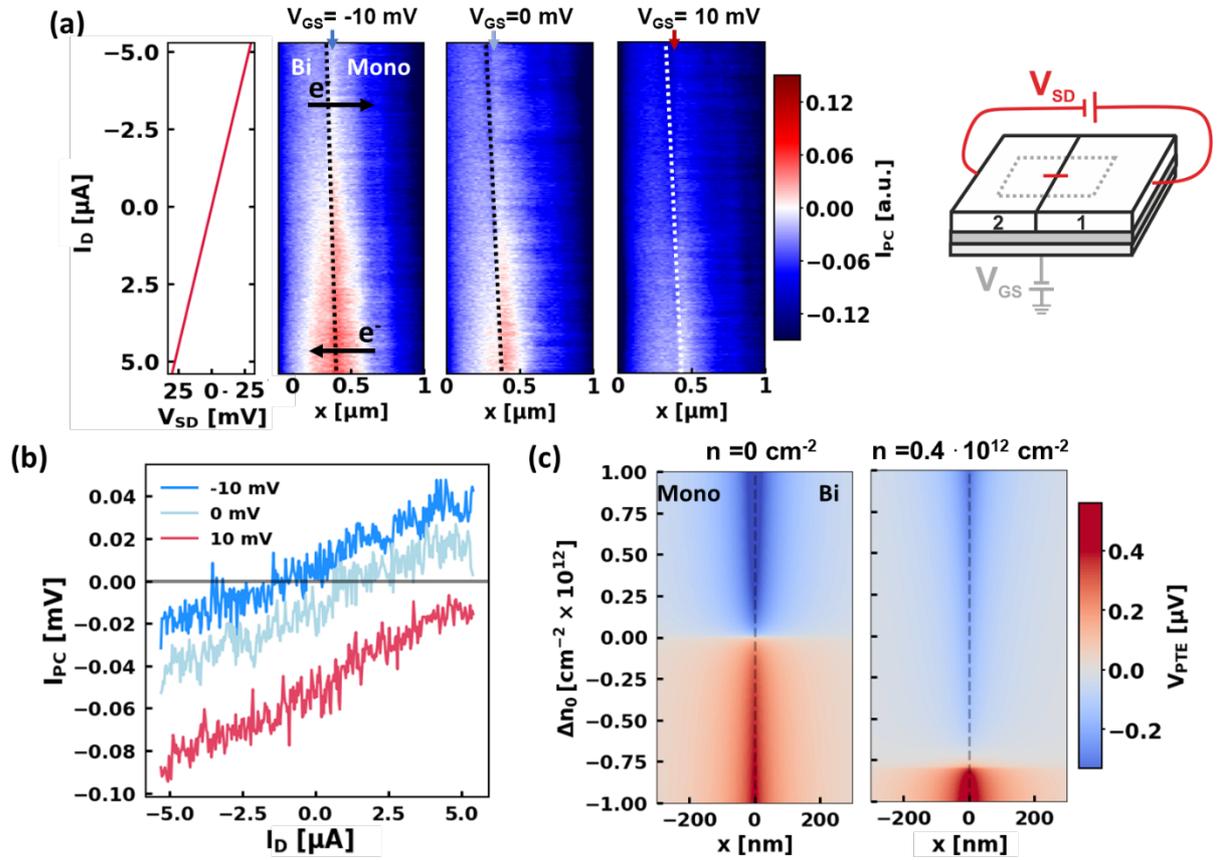

**Figure 4: (a)** 2$^{nd}$ harmonic photocurrent maps measured across the interface, as a function of $I_D$. The data are acquired with $V_{SD}$ ranging from -30 mV to 30 mV in steps of 2 mV, with 10 lines measured for each step. The three maps are measured for two different $V_{GS}$ value around the CNPs, as indicated by the title on top of each map, $V_{GS}$=-10 mV, $V_{GS}$=0 mV and $V_{GS}$=10 mV. The graph to the left of the maps shows $I_{SD}$ with respect to the $V_{SD}$ applied, recorded during the PC map taken at $V_{GS}$= -10 mV, y-axis and color bar are shared. The interface is indicated with the dotted line and the arrows indicate the direction of electron flow respectively at the top and bottom halves of the maps, corresponding to the sign of the applied $V_{SD}$. The schematic on the right indicates that $V_{SD}$ is varied during the acquisition of these maps while $V_{GS}$ is kept constant. **(b)** Vertical line cuts taken at the position in the ML indicated by the arrows in (a). **(c)** Numerically calculated spatial dependence of the photo thermoelectric voltage generated at a ML/BL interface (located at x=0) varying the values of the parameter of $\Delta n_0$ for two different carrier density values: n=0 cm$^{-2}$ in and n=0.4·10$^{12}$ cm$^{-2}$. The spatial coordinate x is perpendicular to the interface and the parameter $\Delta n_0$ mimics the deviation of the carrier density n induced by the $I_{SD}$ - induced LRD around the interface.